\RequirePackage{fix-cm}
\documentclass[twocolumn]{svjour3}          
\smartqed  
\usepackage{booktabs}

\usepackage[math]{blindtext}
\usepackage{mwe}
\usepackage{graphicx}
\usepackage{xcolor}
\usepackage{hyperref}
\usepackage{amsmath}
\usepackage{amssymb}
\usepackage{listings}
\definecolor{mygreen}{rgb}{0,0.6,0}
\definecolor{mygray}{rgb}{0.5,0.5,0.5}
\definecolor{mymauve}{rgb}{0.58,0,0.82}
\definecolor{TUMBlau}{RGB}{0,101,189} 
\definecolor{TUMBlauDunkel}{RGB}{0,82,147} 
\definecolor{TUMBlauHell}{RGB}{152,198,234} 
\definecolor{TUMBlauMittel}{RGB}{100,160,200} 
\definecolor{TUMElfenbein}{RGB}{218,215,203} 
\definecolor{TUMGruen}{RGB}{162,173,0} 
\definecolor{TUMGruenHell}{RGB}{188,207,30} 
\definecolor{TUMGruenDunkel}{RGB}{22,164,00} 
\definecolor{TUMGruenDunkel2}{HTML}{169100} 
\definecolor{TUMRosa}{RGB}{227,130,143} 
\definecolor{TUMRosaHell}{RGB}{242,144,149} 
\definecolor{TUMOrange}{RGB}{243,145,0} 
\definecolor{TUMOrangeHell}{RGB}{247,166,0} 
\definecolor{TUMSenf}{RGB}{202,171,41} 
\definecolor{TUMSenfHell}{RGB}{232,200,55} 
\definecolor{TUMGrau}{gray}{0.6} 
\usepackage{tikz}
\usetikzlibrary{arrows,positioning, automata, calc, fit, shapes.geometric, decorations.pathreplacing, decorations.pathmorphing, decorations.text, spy, shapes}
\usepackage{pgfplots}
\usepackage{algorithmicx}
\usepackage{algorithm}
\usepackage{algc}
\usepackage{algcompatible}
\usepackage{algpseudocode}
\usepackage{subcaption}
\usepackage[htt]{hyphenat}
\usepackage{balance}
\begin{document}

\title{The Duck's Brain}
\subtitle{Training and Inference of Neural Networks in Modern Database Engines}


\author{Maximilian E. Schüle \and Thomas Neumann \and Alfons Kemper}


\institute{Maximilian E. Schüle \at
              University of Bamberg
              \email{maximilian.schuele@uni-bamberg.de}           
           \and
           Thomas Neumann \at
              TUM
              \email{neumann@in.tum.de}           
           \and
           Alfons Kemper \at
              TUM
              \email{neumann@in.tum.de}           
}

\date{Received: \today}
\journalname{arxiv}

\maketitle

\begin{abstract}
Although database systems perform well in data access and manipulation, their relational model hinders data scientists from formulating machine learning algorithms in SQL.
Nevertheless, we argue that modern database systems perform well for machine learning algorithms expressed in relational algebra.
To overcome the barrier of the relational model, this paper shows how to transform data into a relational representation for training neural networks in SQL:
We first describe building blocks for data transformation, model training and inference in SQL-92 and their counterparts using an extended array data type.
Then, we compare the implementation for model training and inference using array data types to the one using a relational representation in SQL-92 only.
The evaluation in terms of runtime and memory consumption proves the suitability of modern database systems for matrix algebra, although specialised array data types perform better than matrices in relational representation.

\end{abstract}

\section{Introduction}
\label{sec:introduction}
Modern database systems generate code to achieve a nearly hard-coded performance.
In pipelined processing, code-generation eliminates interpreted function calls, so that the generated machine code processes data in-place of CPU registers.
Together with modern hardware trends leading to a performance increase of database servers, code-generation allows database systems to take over more complex computations. 
One example for complex computations is the emergence of machine learning~\cite{DBLP:journals/corr/abs-2207-14529} to solve several tasks such as image classification or even replacing database system's components~\cite{DBLP:conf/debs/HeinrichLKB22,DBLP:journals/pvldb/MaltryD22}.
These tasks rarely happen within database systems but in external tools~\cite{DBLP:conf/sigmod/Renz-WielandGKM22,DBLP:conf/edbt/WenigP22} requiring the data to be extracted from database systems~\cite{DBLP:journals/semweb/NathRPH22}.
Thus, current research mostly focuses on eliminating the extraction process~\cite{DBLP:conf/edbt/MayLPMMCG15,DBLP:journals/pvldb/SchuhknechtPHS21,DBLP:journals/pvldb/ButtersteinMSBZ20,DBLP:journals/pvldb/WingerathGR20,DBLP:conf/edbt/StorlK22} and developing systems that combine data management and machine learning~\cite{DBLP:conf/edbt/RaasveldtHMM18}.
In contrast, in this paper, we argue that code generation allows database systems to perform well for machine learning when training neural networks~\cite{DBLP:conf/sigmod/WieseH21} based on matrix algebra in SQL only~\cite{DBLP:conf/icer/MiedemaAF21,DBLP:conf/ht/SalimzadehGHD22,DBLP:conf/sigmod/OlteanuVZ22,DBLP:conf/sigmod/SchuleSBK021,DBLP:conf/btw/SchulePK019}.

\begin{figure}[tbp]
\centering
\begin{tikzpicture}[nodes={draw},xscale=.7]
\sf
\scriptsize
\draw (0,0) node [draw=none] (matrix) {
$\begin{pmatrix}
  a_{1,1} & \cdots & a_{1,n} \\
  \vdots  & \ddots & \vdots  \\
  a_{m,1} & \cdots & a_{m,n}
 \end{pmatrix} $
};
\draw (3,-1.5) node [draw=none] (relational) {
\begin{tabular}{c c c}
i & j & value \\\hline
1 & 1 & $a_{1,1}$ \\
$\dots$ & $\dots$ & $\dots$ \\
1 & n & $a_{1,n}$ \\
$\dots$ & $\dots$ & $\dots$ \\
m & 1 & $a_{m,1}$ \\
$\dots$ & $\dots$ & $\dots$ \\
m & n & $a_{m,n}$
\end{tabular}
};
\draw (-3,-1.5) node [draw=none] (dt) {
\begin{tabular}{c c c c}
rowno. & col1 & $\dots$ & coln \\\hline
1 & $a_{1,1}$ & $\dots$ & $a_{1,n}$ \\
$\dots$  & $\dots$ & $\dots$ & $\dots$  \\
m & $a_{m,1}$ & $\dots$ & $a_{m,n}$
\end{tabular}
};
\draw[->, above] (matrix) -| node[draw=none,right,align=left]{Relational\\Representation} (relational);
\draw[->, above] (matrix) -| node[draw=none,left,align=right]{Tabular\\Representation} (dt);
\end{tikzpicture}
\caption{Tabular and relational representation of matrices in database systems: the latter is used in this study for representing the weights and training neural networks.}
\label{fig:arrayrep}
\end{figure}
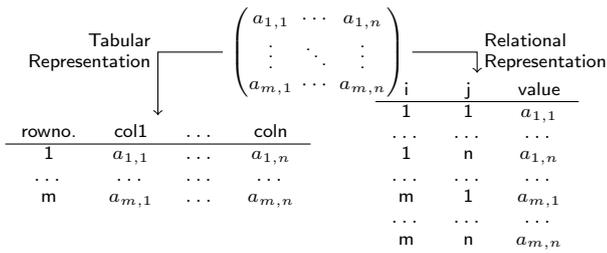

In a previous study, we stated that training neural networks in SQL is possible as long as the database system provides an array data type and recursive tables for gradient descent~\cite{DBLP:conf/ssdbm/SchuleLSK0G21}.
However, the use of an array as a nested data type interferes with the first normal form (referring to the definition of arrays as a non-atomic data type) and requires copying the data between operations.
Instead, to process data in-place of CPU registers, we suggested an array backend for code-generating database systems~\cite{DBLP:conf/ssdbm/SchuleGK021}, which stores matrices in a relational representation (cf. \autoref{fig:arrayrep}). This representation stores arrays in normal form with the indices and the elements as table attributes~\cite{DBLP:conf/edbt/SchuleGK022}.
In a vision paper, Blacher et al.~\cite{DBLP:conf/cidr/BlacherGLKL22} combined our both approaches to show that recursive CTEs (common table expressions)~\cite{DBLP:conf/tapp/Dietrich0G22} can deal with matrices in relational representation as input.
Nevertheless, their study was limited to logistic regression using matrix algebra and no study has benchmarked training neural networks in SQL without further extensions such as arrays before.

In this paper, we even argue that the relational representation allows database systems to efficiently process the computations along with neural networks.
In a preliminary study~\cite{DBLP:conf/btw/SchuleK023}, we demonstrated the ability of database systems to train neural networks using the relational representation.
For training, the performance results were promising for batched processing only and we did not focus on the memory consumption of a relational representation.
Therefore, this paper extends our previous study by model inference, measuring the memory consumption and by including DuckDB~\cite{DBLP:conf/sigmod/RaasveldtM19} as an open-source and modern database engine.
We first describe the mathematical background for reverse mode automatic differentiation that is needed to understand the individual matrix operations (Section~\ref{sec:mlinsql}).
We then discuss the intuitive implementation in Python (Section~\ref{sec:implementationpython}) and deduce an implementation in SQL using the relational representation (Section~\ref{sec:implementationsql92}).
This includes building blocks for data transformation using one-hot-encoding, matrix/Hadamard product and recursive tables to imitate procedural loops.
Afterwards, the paper shows how to transform the relational representation into arrays that are extended for matrix algebra to support model training and inference (Section~\ref{sec:implementationsqlarrays}).
The evaluation (Section~\ref{sec:evaluation}) compares the relational representation to the use of array data types in terms of runtime and memory consumption.
An Python implementation provides the baseline, whose runtime is compared depending on the batch size and the hidden layer size.
Section~\ref{sec:relatedwork} compares our work to existing research on (relational) matrix representations for in-database machine learning.
We conclude with an outlook on optimising recursive tables for this context and on automatically generating the proposed queries (Section~\ref{sec:conclusion}).

\section{Backpropagation for Neural Networks}
\label{sec:mlinsql}
This section first describes the theoretical background for training neural networks and names the variables, which are later used to name the CTEs.
Each variable represents one cached expression computed in the forward pass on function evaluation or in the backward pass on deriving the weight matrices.
To discuss the derivation rules, we exemplary choose a neural network with one hidden layer.
Although this limits the number of hidden layers, the derivation rules can be applied similarly to deep neural networks with further weight matrices in-between.
Thus, the limitation keeps the example short enough to present the implementations in SQL.

\subsection{Gradient Descent in SQL}
\label{sec:sub:autodiff}
A machine learning model can be abstracted as a parameterised model function $m_{w}(x)$.
Its evaluation on data to classify unlabelled data is called \textit{inference}.
\textit{Model training} returns the optimal parameters $w_{\infty}$ minimising a \textit{loss function} $l_{x,y}(w)$, which measures the difference between given labels $y$ and the model $m_{w}(x)$.
Gradient descent minimises this loss function iteratively by moving in the direction of its deepest descent (learning rate $\gamma$), which requires the derivations with respect to each weight (gradient):
$w_{t+1}=\vec{w}_t - \gamma \nabla l_{x,{y}}({w}_t)$, $w_{\infty}\approx \lim\limits_{t \to \infty}{w}_t$.
SQL with recursive CTEs can express the required iterations.
\autoref{code:gdsqlrec} shows gradient descent for simple linear regression:
\begin{align}
   l_{x,y}(a,b)&=(m_{a,b}(x)-y)^2 =(a\cdot x+b-y)^2, \label{eq:gd:loss:salt}\\
   \nabla l_{x,y}(a,b)
  &= \begin{pmatrix}
        \partial l / \partial a \\
        \partial l / \partial b
   \end{pmatrix}
   = \begin{pmatrix}
        2(ax+b-y)\cdot x \\
        2(ax+b-y)
   \end{pmatrix} \label{eq:gd:gradient}.
\end{align}
The base case initialises the weights, each recursion maps to an iteration that updates the weights based on manually derived gradients.
Section \ref{sec:TraininginSQL92} uses this kind of recursion but with the derivations of Section \ref{sec:sub:deriv} for a simple neural network as model.

\begin{lstlisting}[mathescape=true,language=SQL,captionpos=b,caption={Gradient descent in SQL for linear regression: $m_{a,b}(x)=a\cdot x+b \approx y$, five iterations and $\gamma = 0.01$.},label=code:gdsqlrec,float]
create table data (x float, y float);
insert into data ...

with recursive w (id, a, b) as (
 select 0,1::float,1::float
UNION ALL
 select id+1, a-0.01*avg(2*x*(a*x+b-y)),
              b-0.01*avg(2*(a*x+b-y))
 from w, data where id<5 group by id,a,b)
select * from w order by id;
\end{lstlisting}

\subsection{Inference of Neural Networks}
\label{sec:sub:theor}
Neural networks consist of subsequently applied matrix multiplications each followed by an activation function.
They transform an input vector ${x}$ with $m$ attributes into a vector of probabilities for $l$ categories.
With one hidden layer of size $h$, we gain two weight matrices $w_{xh} \in \mathbb{R}^{m\times h}$ and $w_{ho} \in \mathbb{R}^{h \times l}$.
The first one computes the vector $a_{xh} \in \mathbb{R}^h$ for the hidden layer, the second one the result vector $a_{ho} \in \mathbb{R}^l$.
Each activation function returns a normalised value (e.g. $sig(x) \in [0,1]$, \autoref{eq:sig}) that is interpreted as the probability per category.
The result vector is compared to the one-hot-encoded categorical label (${y}_{ones}$).
The difference is elementwisely taken to the power of two ($\Box^{\circ 2}$), which is called mean squared error, a common loss function (\autoref{eq:loss}).
\begin{align}
sig(x) &= (1+e^{-x})^{-1} \label{eq:sig}, \\
m_{w_{xh},w_{ho}}({x}) &= \underbrace{sig(\overbrace{sig ( {x}\cdot w_{xh} )}^{a_{xh}}\cdot w_{ho})}_{a_{ho}} \label{eq:ff}, \\
l(x,y_{ones}) &= (m_{w_{xh},w_{ho}}({x}) - y_{ones})^{\circ 2} \label{eq:loss}.
\end{align}

After computing the loss, reverse mode automatic differentiation computes the derivatives per weight matrix in one pass.
This mode derives a function $f(g(l))$ by decomposing and partially deriving its parts in top-down order: $\frac{\partial f(g(l))}{\partial l} = \frac{\partial f}{\partial g} \cdot \frac{\partial g}{\partial l}$.
Alg.~\ref{alg:autodiff2} shows reverse mode automatic differentiation for matrices~\cite{MLPR}:
The function \texttt{DERIVE} takes as input an arithmetic expression $Z$ and a seed value $seed$ (the parent partial derivation).
The algorithm follows pattern matching on the arithmetic expression $Z$ to compute and further propagate the partial derivatives until arriving at a leaf node.
Upper case letters represent the variable to be parsed, lower case letters represent its actual value evaluated during the foregoing forward pass.
\begin{algorithm}[t]
\caption{Automatic Differentiation (Matrices)}\label{alg:autodiff2}
\begin{algorithmic}[1]
\Function{derive}{$Z,seed$}
   \If {$Z=X+Y$}
   \State \Call{derive}{$X$,$seed$};
   \Call{derive}{$Y$,$seed$}
   \ElsIf {$Z=X\circ Y$}
   \State \Call{derive}{$X$,$seed\circ y$};
   \Call{derive}{$Y$,$seed\circ x$}
   \ElsIf {$Z=X\cdot Y$}
   \State \Call{derive}{$X$,$seed\cdot y^T$};
   \Call{derive}{$Y$,$seed^T\cdot x$}
   \ElsIf {$Z=f(X)$}
   \Call{derive}{$X$,$seed \circ f'(x) $}
   \Else
   $ \frac{\partial}{\partial Z} \gets  \frac{\partial}{\partial Z} + seed$
   \EndIf
\EndFunction
\end{algorithmic}
\end{algorithm}

\subsection{Derivation}
\label{sec:sub:deriv}
\begin{figure}[htbp]
\centering
\begin{tikzpicture}[nodes={draw}, xscale=0.6, yscale=0.8]
\sf
\node (f) at (0,2.05) [rounded rectangle,text=black,minimum height=0.7cm] {
$(sig(sig ( {x}\cdot w_{xh})\cdot w_{ho})-y_{ones})^{\circ 2}$
};
\node (fn)    at (0,0.8) [rounded rectangle,fill=TUMBlauHell,text=black,minimum height=0.7cm] {$\Box ^ {\circ \Box}$};
\node (l)     at (-2.5,0) [rounded rectangle,fill=TUMBlauHell,minimum height=0.7cm] {$-$};
\node (r)     at (2.5,0) [rounded rectangle,minimum height=0.7cm] {$2$};
\node (ll)    at (-5,-0.8) [rounded rectangle,fill=TUMBlauHell,text=black,minimum height=0.7cm] {$sig$};
\node (lr) at (0,-0.8) [rounded rectangle,minimum height=0.7cm] {$y_{ones}$};
\node (cdot2) at (-5,-2.0) [rounded rectangle,fill=TUMBlauHell,text=black,minimum height=0.7cm] {$\cdot$};
\node (woh)   at (-1.5,-2.8) [rounded rectangle,minimum height=0.7cm] {$w_{ho}$};
\node (sig1)  at (-7.5,-2.8) [rounded rectangle,fill=TUMBlauHell,minimum height=0.7cm] {$sig$};
\node (cdot1) at (-7.5,-4.0) [rounded rectangle,fill=TUMBlauHell,minimum height=0.7cm] {$\cdot$};
\node (whx)   at (-4,-4.8) [rounded rectangle,minimum height=0.7cm] {$w_{xh}$};
\node (x)     at (-9,-4.8) [rounded rectangle,minimum height=0.7cm] {$x$};
\draw (fn) edge[->]    (f);
\draw (l)  edge[->]    (fn);
\draw (r)  edge[->]    (fn);
\draw (ll)  edge[bend right,->] node[rounded rectangle,draw=none,black,fill=TUMGruenHell] {$a_{ho}$}  (l);
\draw (lr)  edge[->]  (l);
\draw (x)  edge[->]  (cdot1);
\draw (cdot1)  edge[->] (sig1);
\draw (whx)  edge[->] (cdot1);
\draw (sig1)  edge[bend right,->] node[rounded rectangle,draw=none,black,fill=TUMGruenHell] {$a_{xh}$} (cdot2);
\draw (woh)  edge[->]  (cdot2);
\draw (cdot2)  edge[->] (ll);
\draw (fn) edge[bend right,TUMRosa,<-] (f);
\draw (l)  edge[bend left, TUMRosa,<-] node[rounded rectangle,draw=none,black,fill=TUMRosa] {$l_{ho}$} (fn);
\draw (ll)  edge[bend left, TUMRosa,<-] node[rounded rectangle,draw=none,black,fill=TUMRosa] {$l_{ho}$} (l);
\draw (cdot2)  edge[bend left, TUMRosa,<-]  (ll);
\node[rounded rectangle,draw=none,black,fill=TUMRosa] at ( -5.7,-1.4) {$\delta_{ho}$};
\draw (woh)  edge[bend right, TUMRosa,<-] node[rounded rectangle,draw=none,black,fill=TUMRosa] {$a_{xh}^T\delta_{ho}$} (cdot2);
\draw (sig1)  edge[bend left, TUMRosa,<-] node[rounded rectangle,draw=none,black,fill=TUMRosa] {$l_{xh}$} (cdot2);
\draw (cdot1)  edge[bend left, TUMRosa,<-] (sig1);
\node[rounded rectangle,draw=none,black,fill=TUMRosa] at ( -8.2,-3.4) {$\delta_{xh}$};
\draw (whx)  edge[bend right, TUMRosa,<-] node[rounded rectangle,draw=none,black,fill=TUMRosa] {$x^T \cdot \delta_{xh}$} (cdot1);
\end{tikzpicture}
\caption{Automatic differentiation for $(m_{w_{xh},w_{ho}}({x})-y_{ones})^{\circ 2}$.}
\label{fig:autodiff_nn}
\end{figure}
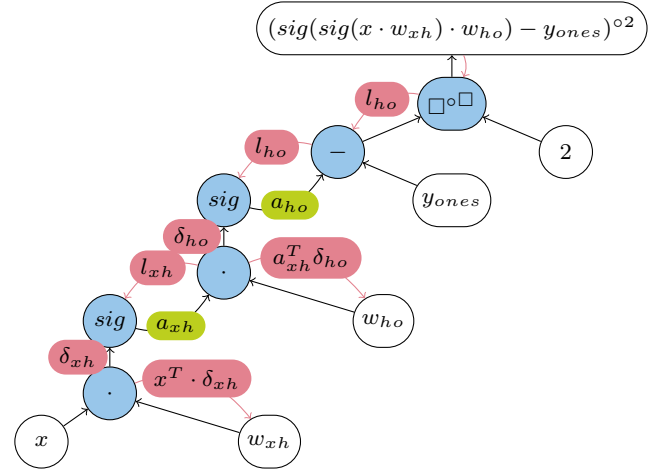

By step-wise applying the derivation rules, we obtain the expression tree shown in \autoref{fig:autodiff_nn}.
The derivative of mean squared error calculates the difference between propagated probabilities and the one-hot-encoded labels (\autoref{eq:out_err}).
This value gets propagated as initial seed value.
Each seed value is elementwise multiplied to each partial derivation, so either the derivation of each activation function (\autoref{eq:d_ho},~\ref{eq:d_xh}) or the matrix multiplication (\autoref{eq:hid_err}).
Finally, the derivation of each weight matrix times the learning rate $\gamma$ is subtracted from the weight matrix to form the updated weights (\autoref{eq:w_ho},~\ref{eq:w_xh}).
\begin{align}
{l}_{ho}  &= 2 \cdot (m_{w_{xh},w_{ho}}({x}) - {y}_{ones}), \label{eq:out_err} \\ 
{\delta}_{ho} &= {l}_{ho} \circ sig'({a}_{ho}) = {l}_{ho} \circ {a}_{ho} \circ (1-{a}_{ho}), \label{eq:d_ho}\\
{l}_{xh}  &= {\delta}_{ho} \cdot w_{ho}^T, \label{eq:hid_err} \\
{\delta}_{xh} &= {l}_{xh} \circ sig'({a}_{xh}) = {l}_{xh} \circ {a}_{xh} \circ (1-{a}_{xh}), \label{eq:d_xh} \\
w_{ho}'     &= w_{ho} - \gamma \cdot  {a}_{xh}^T\cdot {\delta}_{ho}, \label{eq:w_ho} \\
w_{xh}'     &= w_{xh} - \gamma \cdot  {x}^T     \cdot {\delta}_{xh}  \label{eq:w_xh}.
\end{align}

\section{Implementation in Python}
\label{sec:implementationpython}
Having defined the equations for training a neural network, we can deduce a Python implementation (\autoref{fig:nnpy}) that uses NumPy for data loading (line~\ref{lst:nnpy:3}), transformation (lines \ref{lst:nnpy:4}-\ref{lst:nnpy:8}) and generating randomised weights (lines \ref{lst:nnpy:10}-\ref{lst:nnpy:12}).
Afterwards, a procedural loop (line \ref{lst:nnpy:14}) performs gradient descent that updates the weights according to the derivation rules in each iteration (lines \ref{lst:nnpy:15}-\ref{lst:nnpy:24}).
So each variable represents one equation needed to backpropagate the loss.

\section{Implementation in SQL-92}
\label{sec:implementationsql92}
In order to update the weight matrices of neural networks in SQL, we need to map matrix multiplication ($X\cdot Y$), function application ($f(X)$) and elementwise operations (addition:~$X+Y$, Hadamard multiplication $X\circ Y$) to the relational representation in SQL.
For binary elementwise operations such as Hadamard multiplication or addition/subtraction, a join on the indices combines both tables so that the arithmetic operation is part of the select-clause.
Multiplication of two matrices $m \in \mathbb{R}^{m \times o}$ and $n \in \mathbb{R}^{o \times n}$ with equal inner dimensions is defined as the sum of the product over $o$ row/column elements for each entry $ (m \cdot n)_{ij} = \sum_{k=1}^{o} m_{ik}n_{kj}$.
In relational algebra, this means a join on the inner index, followed by a summation:
$\gamma_{m.i, n.j, sum(m.v\cdot n.v)}(m\bowtie_{m.j = n.i}n)$.
To transpose a matrix in relational representation, only the indices have to be renamed.
The corresponding SQL building blocks are shown in \autoref{fig:mmul} with their NumPy counterparts in \autoref{fig:numpymul}.

\begin{lstlisting}[mathescape=true,escapechar=|,language=Python,captionpos=b,caption={Training a neural network with NumPy.},label=fig:nnpy,float]
import numpy as np
# load data
arr = np.loadtxt("iris.csv", delimiter=",", dtype=float,skiprows=1)|\label{lst:nnpy:3}|
X = arr[:,0:4]/10|\label{lst:nnpy:4}|
y = arr[:,4].astype(int)
# one-hot-encode y
y_oh = np.zeros((y.size, y.max()+1))
y_oh[np.arange(y.size),y] = 1 # one-hot-encode|\label{lst:nnpy:8}|
# initialise weights
np.random.seed(1)|\label{lst:nnpy:10}|
w_xh = 2*np.random.random((X[0].size,20)) - 1
w_ho = 2*np.random.random((20,3)) - 1        |\label{lst:nnpy:12}|
# train
for j in range(10):|\label{lst:nnpy:14}|
    print("Iteration: " + str(j))|\label{lst:nnpy:15}|
    # sigmoid(x*w\_xh)
    a_xh = 1/(1+np.exp(-np.dot(X,w_xh)))
    # sigmoid(a\_xh*w\_ho)
    a_ho = 1/(1+np.exp(-np.dot(a_xh,w_ho)))
    l_ho = 2*(a_ho - y_oh)
    print("Loss: " + str(np.mean(np.abs(l_ho))))
    d_ho = l_ho * a_ho * (1-a_ho)
    l_xh = d_ho.dot(w_ho.T)
    d_xh = l_xh * a_xh * (1-a_xh)
    w_ho -= 0.01 * a_xh.T.dot(d_ho)
    w_xh -= 0.01 * X.T.dot(d_xh)|\label{lst:nnpy:24}|
\end{lstlisting}

\begin{lstlisting}[mathescape=false,mathescape=true,escapechar=|language=Python,captionpos=b,caption={Building blocks for matrices in NumPy.},label=fig:numpymul,float]
m.dot(n)           # matrix multiplication
m * n              # hadamard multiplication
1/(1+np.exp(-m))   # sigmoid function
m.T                # transpose
\end{lstlisting}
\begin{lstlisting}[mathescape=false,mathescape=true,escapechar=|language=SQL,captionpos=b,caption={Building blocks for matrices in SQL-92.},label=fig:mmul,float]
-- create two matrices m and n
create table m (i int, j int, v float); create table n (i int, j int, v float);
insert into m ...
-- matrix multiplication
select m.i, n.j, SUM(m.v*n.v))
from m inner join n on m.j=n.i group by m.i, n.j
-- hadamard multiplication
select m.i, m.j, m.v*n.v
from m inner join n on m.i=n.i and m.j=n.j
-- sigmoid function
select i, j,  1/(1+exp(-v)) from m;
-- transpose
select i as j, j as i, v from m;
\end{lstlisting}

\subsection{Transformation into Relational Representation}
\label{sec:TransformationintoRelationalRepresentation}
To train the neural network in SQL, we first have to convert the data into the relational representation (\autoref{fig:onehotsql92}).
Therefore, we create a table of two indices and a value corresponding to the two-dimensional feature matrix (\texttt{img}: $\{[i,j,v]\}$, line \ref{lst:onehotsql92:3}).
We assign a column index $j$ to each attribute of the original input table (lines \ref{lst:onehotsql92:5}-\ref{lst:onehotsql92:8}) and use the row number as index $i$.
Afterwards, we one-hot-encode the label:
We generate a sparse matrix containing only the one values (line \ref{lst:onehotsql92:11}) and a matrix shape---defined by all indices within the dimensions---out of \texttt{null} values (lines \ref{lst:onehotsql92:12}-\ref{lst:onehotsql92:14}).
Then, an outer join (lines \ref{lst:onehotsql92:11}/\ref{lst:onehotsql92:15}) combines both tables and assigns zero to missing values (\texttt{coalesce}: line \ref{lst:onehotsql92:10}).

\begin{figure*}[tbp]
\centering
\begin{tikzpicture}[nodes={draw},xscale=1.3]
\sf
\scriptsize
\draw (2.25,-1.5) node [draw=none] (label) {
\begin{tabular}{c c c}
i & j & v \\\hline
1 & 1 & $1$ \\
1 & 2 & $0$ \\
1 & 3 & $0$ \\
$\dots$  & $\dots$ & $\dots$ \\
150 & 3 & $1$ \\
\end{tabular}
};
\draw (4.7,-1.5) node [draw=none] (feature) {
\begin{tabular}{c c c}
i & j & v \\\hline
1 & 1 & $5.1$ \\
1 & 2 & $3.5$ \\
1 & 3 & $1.4$ \\
1 & 4 & $0.2$ \\
$\dots$  & $\dots$ & $\dots$ 
\end{tabular}
};
\draw (-2.75,-1.5) node [draw=none] (dt) {
\begin{tabular}{c c c c c c}
row & sepal length & s. width & petal length & p. width & species \\\hline
1 & 5.1&3.5&1.4&0.2&0 \\
$\dots$  & $\dots$ & $\dots$ & $\dots$ & $\dots$ & $\dots$  \\
$\dots$  & $\dots$ & $\dots$ & $\dots$ & $\dots$ & $\dots$  \\
$\dots$  & $\dots$ & $\dots$ & $\dots$ & $\dots$ & $\dots$  \\
150 & 5.9&3.0&5.1&1.8&2
\end{tabular}
};
\draw[->, above] (-3,-0.5) -| (-3,0) -| node[draw=none,right,align=left]{Feature Matrix} (feature);
\draw[-] (-5.8,-0.5) -- (-0.65,-0.5);
\draw[->, above] (-0.1,-0.5) -| (-0.1,-0.25) -| node[draw=none,right,align=left]{One-Hot-Encoded} (label);
\end{tikzpicture}
\caption{Transformation of the original data set into the relational representation.}
\label{fig:transform}
\end{figure*}
\begin{lstlisting}[mathescape=true,escapechar=|,language=SQL,captionpos=b,caption={Data transformation: Feature matrix \texttt{img} and one-hot-encoded label \texttt{one\_hot}.},label=fig:onehotsql92,float]
create table if not exists iris (id serial, sepal_length float, sepal_width float, petal_length float, petal_width float, species int);
copy iris from './iris.csv' delimiter ',' HEADER CSV;
create table img (i int, j int, v float);|\label{lst:onehotsql92:3}|
create table one_hot(i int, j int, v int);
insert into img (|\label{lst:onehotsql92:5}|
  select id,1,sepal_length/10 from iris);
insert into img (
  select id,2,sepal_width/10 from iris);
insert into img (
  select id,3,petal_length/10 from iris);
insert into img (
  select id,4,petal_width/10 from iris);|\label{lst:onehotsql92:8}|
insert into one_hot(
 select n.i, n.j, coalesce(i.v,0), i.v|\label{lst:onehotsql92:10}|
 from (select id,species+1 as species,1 as v|\label{lst:onehotsql92:11}|
       from iris) i right outer join
  (select a.i, b.j|\label{lst:onehotsql92:12}|
   from (select generate_series as i
         from generate_series(1,select count(*) from iris)) a,
        (select generate_series as j
         from generate_series(1,4)) b|\label{lst:onehotsql92:14}|
 ) n on n.i=i.id and n.j=i.species order by i,j);|\label{lst:onehotsql92:15}|
\end{lstlisting}

\subsection{Training in SQL-92}
\label{sec:TraininginSQL92}
After transforming the data, we can create and initialise the weights again in relational representation.
Using \texttt{generate\_series} according to the matrix dimensions together with \texttt{random}, we initialise all required weight matrices.
\begin{lstlisting}[mathescape=false,mathescape=true,escapechar=|language=SQL,captionpos=b,caption={Create and initialise weights in SQL-92.},label=fig:weights,float]
create table w_xh (i int, j int, v float);
create table w_ho (i int, j int, v float);
insert into w_xh (select i.*,j.*,random()*2-1
                  from generate_series(1,4) i,
                       generate_series(1,20) j);
insert into w_ho (select i.*,j.*,random()*2-1
                  from generate_series(1,20) i,
                       generate_series(1,3) j);
\end{lstlisting}

The feature matrix in relational representation forms the input for training the neural network within a recursive CTE (\autoref{fig:nnsql92}) that computes the weights per iteration of gradient descent.
As we need to compute all weights within the recursive CTE, a unique number (\texttt{id}) identifies each weight matrix.
Thus a union of all weight matrices forms the base case for the recursion.
Within the recursive step, nested CTEs help to evaluate the model (lines \ref{lst:nnsql92:6}-\ref{lst:nnsql92:15}), to backpropagate the loss (lines \ref{lst:nnsql92:16}-\ref{lst:nnsql92:29}) and to compute the derivative per weight matrix (lines \ref{lst:nnsql92:30}-\ref{lst:nnsql92:37}).
The first CTE \texttt{w\_}---just referring to the original weights---is necessary, as PostgreSQL only allows one reference to the recursive table.
Each following CTE computes one matrix operation, so either a matrix or a Hadamard multiplication, whose CTE name refers to the variable name (cf.~Section~\ref{sec:sub:theor}).
Finally, the weights were updated by subtracting their derivatives (lines \ref{lst:nnsql92:39}-\ref{lst:nnsql92:41}).

\begin{lstlisting}[mathescape=true,escapechar=|,language=SQL,captionpos=b,caption={Training a neural network in SQL-92.},label=fig:nnsql92,float]
with recursive w (iter,id,i,j,v) as (
 (select 0,0,* from w_xh union
  select 0,1,* from w_ho)
 union all
 ( with w_ as (
  -- recursive reference only allowed once in PSQL
   select * from w
  ), a_xh(i,j,v) as (|\label{lst:nnsql92:6}| -- sig(img * w\_xh)
   select m.i, n.j, 1/(1+exp(-SUM(m.v*n.v)))
   from img as m inner join w_ as n on m.j=n.i
   where n.id=0 and
     n.iter=(select max(iter) from w_) -- w\_xh
   group by m.i, n.j
  ), a_ho(i,j,v) as ( -- sig(a\_xh * w\_ho)
   select m.i, n.j, 1/(1+exp(-SUM(m.v*n.v)))
   from a_xh as m inner join w_ as n on m.j=n.i
   where n.id=1 and
     n.iter=(select max(iter) from w_)  -- w\_ho
   group by m.i, n.j|\label{lst:nnsql92:15}|
  ), l_ho(i,j,v) as (|\label{lst:nnsql92:16}| -- 2 * (a\_ho-y\_ones)
   select m.i, m.j, 2*(m.v-n.v)
   from a_ho as m inner join one_hot as n
        on m.i=n.i and m.j=n.j
  ), d_ho(i,j,v) as ( -- l\_ho ° a\_ho ° (1-a\_ho)
   select m.i, m.j, m.v*n.v*(1-n.v)
   from l_ho as m inner join a_ho as n
        on m.i=n.i and m.j=n.j
  ), l_xh(i,j,v) as ( -- d\_ho * w\_ho\^~T
   select m.i, n.i as j, SUM (m.v*n.v)
   from d_ho as m inner join w_ as n on m.j=n.j
   where n.id=1 and
     n.iter=(select max(iter) from w_)  -- w\_ho
   group by m.i, n.i|\label{lst:nnsql92:29}|
  ), d_xh(i,j,v) as ( -- l\_xh ° a\_xh ° (1-a\_ho)|\label{lst:nnsql92:30}|
   select m.i, m.j, m.v*n.v*(1-n.v)
   from l_xh as m inner join a_xh as n
        on m.i=n.i and m.j=n.j
  ), d_w(id,i,j,v) as (
   select 0, m.j as i, n.j, SUM (m.v*n.v)
   from img as m inner join d_xh as n on m.i=n.i
   group by m.j, n.j
   union
   select 1, m.j as i, n.j, SUM (m.v*n.v)
   from a_xh as m inner join d_ho as n on m.i=n.i
   group by m.j, n.j|\label{lst:nnsql92:37}|
  )
 select iter+1, w.id, w.i, w.j, w.v-0.01*d_w.v|\label{lst:nnsql92:39}|
 from w_ as w, d_w
 where iter < 20 and w.id=d_w.id and w.i=d_w.i and w.j=d_w.j|\label{lst:nnsql92:41}|
 )
) select * from w;
\end{lstlisting}

\subsection{Inference in SQL-92}
In order to predict the accuracy of the trained weights, an SQL query measures the number of correctly classified labels (\autoref{fig:predictionsql}).
Evaluating the model (lines \ref{lst:predictionsql:3}-\ref{lst:predictionsql:9}) returns a vector of probabilities per tuple and category.
The SQL query ranks the predicted probabilities per tuple (line \ref{lst:predictionsql:2}) and the one-hot-encoded vector of the original labels (line \ref{lst:predictionsql:11}) to compare whether the index of the highest probability matches the index of the one value (line \ref{lst:predictionsql:14}).
Although window functions were used for the ranking, they could be replaced by an anti-join using \texttt{not exists} to conform SQL-92.

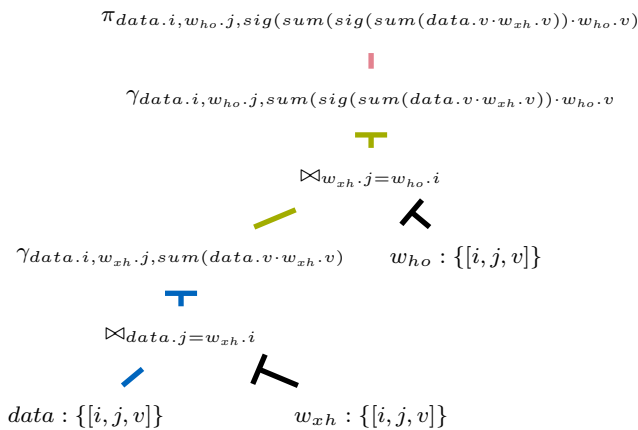
\begin{figure}[htbp]
\centering
\begin{tikzpicture}[-,>=stealth',shorten >=1pt,auto,xscale=2.5,yscale=1.4]
\sf\small
  \tikzstyle{every state}=[draw=none,text=black,rectangle]

  \node[state]		(d)   at (-1.5,-5.25)	{$data: \{[i,j,v]\}$};
  \node[state]		(wxh)   at (-0,-5.25)	{$w_{xh}: \{[i,j,v]\}$};
  \node[state]		(join1)  at (-1.0,-4.5)	{$\bowtie_{data.j = w_{xh}.i}$};
  \node[state]		(agg1)at (-1.0,-3.75) 	{$\gamma_{data.i, w_{xh}.j, sum(data.v\cdot w_{xh}.v)}$};
  \node[state]		(join2)   at (0,-3)	{$\bowtie_{ w_{xh}.j = w_{ho}.i }$};
  \node[state]		(who)   at (0.5,-3.75 )	{$w_{ho}: \{[i,j,v]\}$};
  \node[state]		(agg2)  at (0,-2.25) 	{$\gamma_{data.i, w_{ho}.j, sum(sig(sum(data.v\cdot w_{xh}.v)) \cdot  w_{ho}.v }$};
  \node[state]		(sig)   at (0,-1.5)	{$\pi_{data.i, w_{ho}.j, sig(sum(sig(sum(data.v\cdot w_{xh}.v)) \cdot  w_{ho}.v)}$};

  \path
        (d)	edge[TUMBlau,line width=2pt]	node (node1) {} (join1)
        (wxh)	edge[-|,line width=2pt]		node (node2) {} (join1)
        (join1)	edge[-|,TUMBlau,line width=2pt]	node (node3) {} (agg1)
        (agg1)	edge[TUMGruen,line width=2pt]	node (node4) {} (join2)
        (who)	edge[-|,line width=2pt]		node (node5) {} (join2)
        (join2)	edge[-|,TUMGruen,line width=2pt]node (node6) {} (agg2)
        (agg2)	edge[TUMRosa,line width=2pt]	node (node7) {} (sig)
	    ;
\end{tikzpicture}
\caption{Query plan for model inference in relational representation: Five pipelines, one for materialising each weight matrix (black), one for each matrix multiplication ({\color{TUMBlau}blue}, {\color{TUMGruen}green}), one for the output ({\color{TUMRosa}red}).}
\label{fig:queryplan}
\end{figure}
\begin{lstlisting}[mathescape=true,escapechar=|,language=SQL,captionpos=b,caption={Prediction in SQL:2003 (with window functions).},label=fig:predictionsql,float]
select iter, count(*)::float/(
   select count(distinct i) from one_hot)
from (
 select *, rank() over
    (partition by m.i,iter order by v desc)|\label{lst:predictionsql:2}|
 from (
  select m.i,n.j,1/(1+exp(-sum(m.v*n.v))) as v,m.iter|\label{lst:predictionsql:3}|
  from (
   select m.i,n.j,1/(1+exp(-sum(m.v*n.v))) as v,iter
   from img AS m inner join w as n on m.j=n.i
   where n.id=0
   group by m.i, n.j, iter ) AS m
  inner join w as n on m.j=n.i
  where n.id=1 and n.iter=m.iter
  group by m.i, n.j, m.iter|\label{lst:predictionsql:9}|
 ) m ) pred,
 (select *, rank() over (partition by m.i order by v desc) from one_hot m) test|\label{lst:predictionsql:11}|
where pred.i=test.i and pred.rank = 1
      and test.rank=1
group by iter, pred.j=test.j
having (pred.j=test.j)=true|\label{lst:predictionsql:14}|
order by iter
\end{lstlisting}

\begin{lstlisting}[mathescape=true,escapechar=|,language=SQL,captionpos=b,caption={Transformation of a relational representation into an array representation.},label=lst:arraytransfrom,float]
create table weights ( |\label{lst:arraytransfrom:1}|
   w_xh float[][], w_ho float[][]) |\label{lst:arraytransfrom:2}|
insert into weights (select
 (select array_agg(js order by i) from ( |\label{lst:arraytransfrom:3}|
   select array_agg(v order by j) as js |\label{lst:arraytransfrom:4}|
   from w_xh group by i) tmp),
 (select array_agg(js order by i) from ( |\label{lst:arraytransfrom:5}|
   select array_agg(v order by j) as js |\label{lst:arraytransfrom:6}|
   from w_ho group by i) tmp));
\end{lstlisting}

\autoref{fig:queryplan} shows the query plan for model inference (without prediction accuracy) in a relational representation.
Each weight matrix is materialised for the equi-join, so the pipeline for the data can flow until reaching an aggregation.
Each matrix multiplication consists of an aggregation and thus forms a full pipeline breaker.
As we will later see in the evaluation, this limits the number of possibly processed tuples for model inference by the memory required to materialise the data within each aggregation.
A solution to overcome materialisation for an aggregation would be another physical implementation:
Instead of hashing the values for aggregation, sorted input (ordered by the index for the row number \texttt{i}) would allow for continuous output.
This would work well for data stream management systems and when the result table can be stored to disk.

\section{Implementation in SQL with Arrays}
\label{sec:implementationsqlarrays}
The presence of an extended array data type that supports matrix algebra allows SQL to train models without the need for one CTE representing each matrix.
In a previous study~\cite{DBLP:conf/btw/Schule23}, we compared the performance of gradient descent as an operator to using recursive tables.
In contrast, this section focuses on the transformation of a relational matrix into its array format, its usage for training within a recursive table and inference as later used in the evaluation for comparison with the relational representation.

\subsection{Transformation into Array Representation}
Transforming the relational representation into the array representation means reconstructing the matrix based on its indices created in Section~\ref{sec:TransformationintoRelationalRepresentation}.
\autoref{lst:arraytransfrom} shows this transformation by first creating a table for the weight matrices with one attribute (\texttt{float[][]}) per weight matrix (line~\ref{lst:arraytransfrom:1}-\ref{lst:arraytransfrom:2}).
Afterwards, the matrix in relational representation can be constructed for which \texttt{array\_agg} as aggregation function is used:
First for the inner dimension (\texttt{j}), we group by the higher dimension (\texttt{i}) and---within the aggregation---sort by the current dimension that is condensed (\texttt{j}) (line~\ref{lst:arraytransfrom:4},\ref{lst:arraytransfrom:6}).
Finally for the last dimension (\texttt{i}) we can group the matrix together without any aggregation key (line~\ref{lst:arraytransfrom:3},\ref{lst:arraytransfrom:5}).

\subsection{Training}
A recursive table representing the weights is used for model training using gradient descent similar to Section~\ref{sec:TraininginSQL92}.
The base case takes the initial weights as input and assigns them the number 0, marking them as the initial weights for the first iteration (\autoref{code:nntrainarray}, line~\ref{lst:trainarray:1}).
The recursive case evaluates the model function first (line~\ref{lst:trainarray:7}-\ref{lst:trainarray:8}), computes the loss (line~\ref{lst:trainarray:6}) and applies the backpropagation rules (line~\ref{lst:trainarray:4}-\ref{lst:trainarray:5}) to finally update the weights per iteration (line~\ref{lst:trainarray:2}-\ref{lst:trainarray:3}).
The required extensions to an array data type are elementwise operations (\texttt{-}: subtraction, \texttt{*}: Hadamard multiplication $\circ$), matrix multiplication (\texttt{**}), elementwise function application like a map function (here hard-coded for \texttt{sig}), transposition, and elementwise aggregation (\texttt{sum}).

\begin{lstlisting}[mathescape=true,escapechar=|,language=SQL,captionpos=b,caption={Backpropagation for a neural network using a recursive table and an array data type.},label=code:nntrainarray,float]
with recursive w (id,w_xh,w_ho) as (
 select 0, w_xh, w_ho from weights|\label{lst:trainarray:1}|
union all
 select id+1,|\label{lst:trainarray:2}|
        w_xh - 0.01 * sum(transpose(img)*d_xh),
        w_ho - 0.01 * sum(transpose(a_xh)*d_ho)|\label{lst:trainarray:3}|
 from ( select l_xh**(a_xh**(1-a_xh)) as d_xh, *|\label{lst:trainarray:4}|
  from ( select d_ho*transpose(w_ho) as l_xh, *
   from (
    select l_ho**(a_ho**(1-a_ho)) as d_ho, *|\label{lst:trainarray:5}|
    from ( select 2*(a_ho-one_hot) as l_ho, *|\label{lst:trainarray:6}|
     from ( select sig(a_xh*w_ho) as a_ho, *|\label{lst:trainarray:7}|
      from ( select sig(img*w_xh) as a_xh, *
       from ( select * from data),w
        where id < 20))))))|\label{lst:trainarray:8}|
 group by id, w_ho, w_xh
) select * from w:
\end{lstlisting}
\subsection{Inference}
For model inference, the model function has to be evaluated similarly as for training but with the optimal weights.
In order to calculate the prediction accuracy defined as the number of correctly classified labels, the output vector of probabilities has to be compared to the one-hot-encoded origin label of a test data set.
So the index of the highest probability should match the position of the one in one-hot-encoding, for which we created an array function \texttt{highestposition} that returns the index of the maximum value (e.g., \texttt{highestposition([0,0.6,0.4])=1}).
To calculate the accuracy, we count the number of correctly and incorrectly classified labels (\autoref{fig:inferencesqlarrays}, line~\ref{lst:inferencearray:1}-\ref{lst:inferencearray:3}) and divide the number of correct labels through the number of all (line~\ref{lst:inferencearray:4}-\ref{lst:inferencearray:5}).
\begin{lstlisting}[mathescape=true,escapechar=|,language=SQL,captionpos=b,caption={Accuracy evaluation on predicted probabilities with an array data type.},label=fig:inferencesqlarrays,float]
with test as (
 select correct, count(*) as cnt from (|\label{lst:inferencearray:1}|
  select highestposition(sig(sig(img**w_xh)**w_ho))=highestposition(one_hot) as correct|\label{lst:inferencearray:2}|
  from data,weights)
 group by correct)|\label{lst:inferencearray:3}|
select cnt*1.0/(select sum(cnt) from test t2)|\label{lst:inferencearray:4}|
from test t1 where correct=true;|\label{lst:inferencearray:5}|
\end{lstlisting}

\section{Evaluation}
\label{sec:evaluation}
\textit{System:} Ubuntu 22.04 LTS, Intel(R) Xeon(R) W-2295 CPU (18 cores @ 3.00GHz supporting hyper-threading), 128~GB~DDR4~RAM.

We compare the performance of the relational representation for matrices (\textit{SQL-92}, Section~\ref{sec:implementationsql92}) to their representation as an array data type (\textit{SQL + Arrays}, Section~\ref{sec:implementationsqlarrays}).
We apply both representations for use within neural networks in SQL and let the benchmarks\footnote{\url{https://gitlab.db.in.tum.de/MaxEmanuel/nn2sql}} run in Umbra~\cite{DBLP:conf/cidr/NeumannF20}, PostgreSQL (PSQL) 14.5~\cite{DBLP:conf/sigmod/StonebrakerR86} and DuckDB 0.8.1~\cite{DBLP:conf/sigmod/RaasveldtM19} as target engines.
The implementation with NumPy (\autoref{fig:nnpy}, Section~\ref{sec:implementationpython}) serves as the baseline.
We use two different data sets: Fisher's Iris flower data~\cite{fisher1936use} (four attributes, one label) and the MNIST data~\cite{DBLP:conf/cvpr/CiresanMS12} for image classification (ten categories, 784 pixels, excerpt of 6000 tuples).

\subsection{Relational Array Representation}
\label{sec:relarray}
Before benchmarking composed matrix operations for neural networks, we want to focus on the memory impact of a single multiplication.
The advantages of a matrix data type are the less integration effort as arrays are supported by most of the database systems and the interaction with lambda functions as provided in DuckDB (\texttt{list\_transform}).
Although extending an array data type by matrix operations comes with these advantages, an array data type nonetheless violates the first normal form and relational database systems support a relational representation natively.
Beyond a matrix data type, we want to investigate the suitability of a relational matrix representation within database systems.

\begin{figure}[tbp]
\begin{center}
\includegraphics[width=0.45\textwidth]{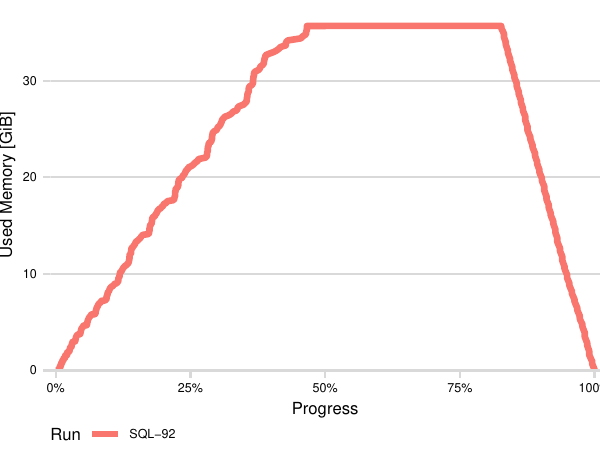}
\includegraphics[width=0.45\textwidth]{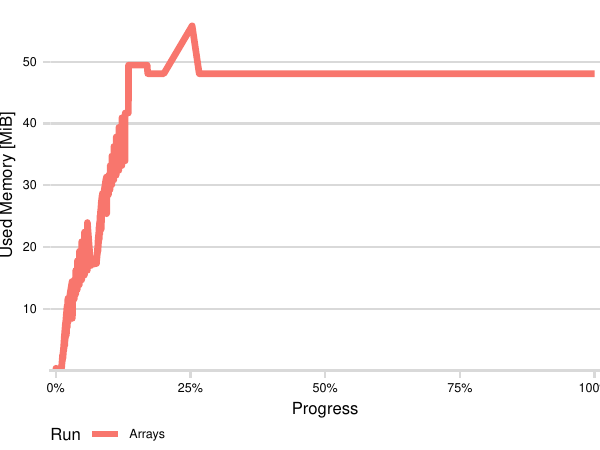}
\end{center}
\caption{Memory usage of a matrix multiplication $M \cdot N, M,N \in \mathbb{R}^{1000\times 1000} $ either in relational representation (\texttt{SQL-92}) or using arrays (\texttt{Arrays}).}
\label{fig:mem}
\end{figure}

\autoref{fig:mem} shows the memory consumption of a matrix multiplication using a relational representation compared to arrays:
Each matrix has a size of 8~MB (assuming 8 B per double-precision floating point number), thus all three matrices have a bare size of 24~MB.
Its relational counterpart requires three times its size with 8 Byte for each dimension.
When joining the tables to compute the matrix multiplication (cf.~\autoref{lst:sqlmatrix}), the size of the intermediate result is a thousandfold (1000 tuples per entry) before aggregation.
Thus a matrix representation---although fast---wastes space on multiplication.

\begin{lstlisting}[mathescape=false,language=SQL,captionpos=b,caption={Matrix multiplication in SQL.},label=lst:sqlmatrix,float=tbp]
create table a (i int, j int, v float);
create table b (i int, j int, v float); /*...*/
select a.i,b.j, sum(a.v*b.v) from a,b
where a.j=b.i group by a.i,b.j;
\end{lstlisting}

\begin{figure}[tbp]
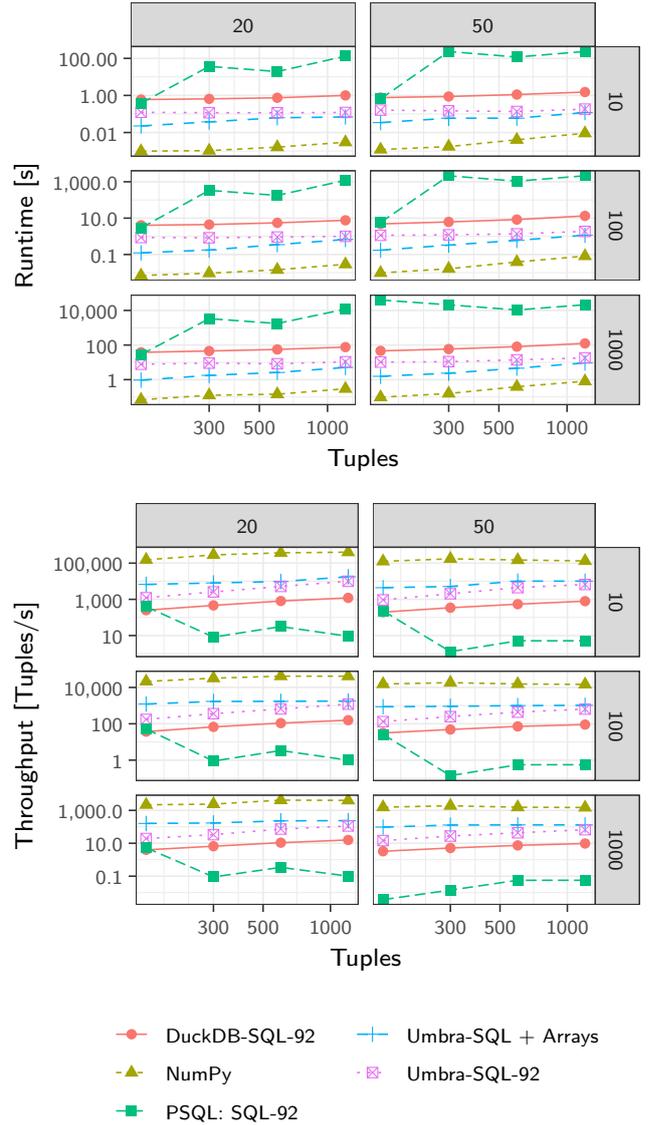

\sf
{
   \input{img/iris.tex}
   \vspace{-1cm}
   \input{img/iristhroughput.tex}
}
   \caption{Performance: Training a neural network with one hidden layer (size 20/50, 10/100/1000 iteration).}
   \label{fig:eval:iris}
\end{figure}

\begin{figure}[tbp]
\sf
   \includegraphics[width=\linewidth]{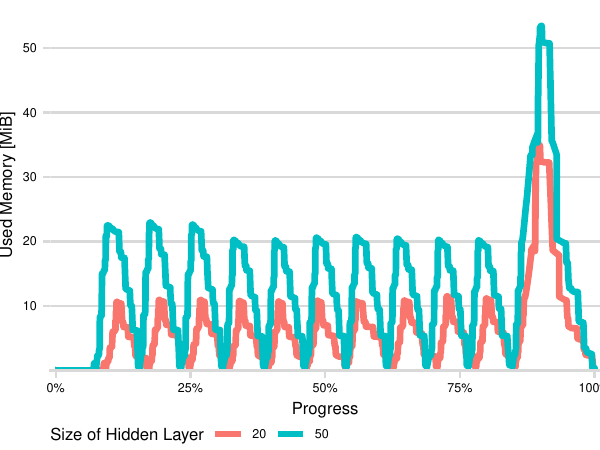}
   \caption{SQL-92: Memory consumption (10 iterations, batch size = 150 Tuples, hidden layer size 20/50).}
   \label{fig:eval:irissql9210150}
\end{figure}
\begin{figure}[tbp]
\sf
   \includegraphics[width=\linewidth]{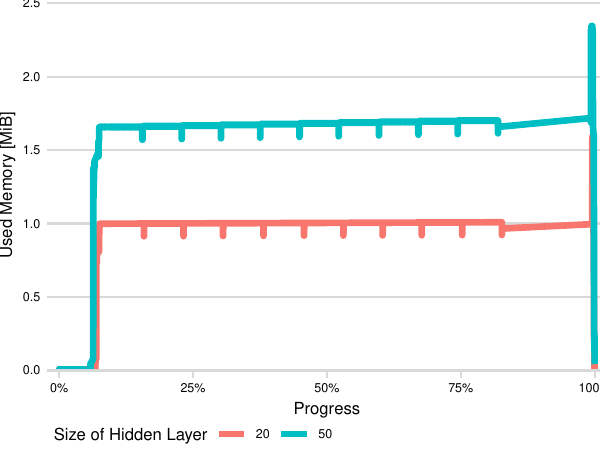}
   \caption{SQL + Arrays: Memory consumption (10 iterations, batch size = 150 Tuples, hidden layer size 20/50).}
   \label{fig:eval:irisarray10150}
\end{figure}

\subsection{Scaling the Number of Input Tuples}
\autoref{fig:eval:iris} shows the performance in terms of runtime and throughput using the Iris data set.
As we are interested in the performance numbers and not in the model quality, we replicate the Iris flower data set for the first benchmark to enable a flexible input size.
A neural network with one hidden layer is trained to classify the flower category.
We vary the size of the training data set, the number of iterations and the size of the hidden layer. 
\subsubsection{Throughput}
Although the NumPy implementation outperforms both SQL variants, the performance increase of Umbra with its in-memory performance in comparison to PostgreSQL is visible.
The performance of DuckDB, as a vectorised engine without code-compilation, lies in-between: faster than PostgreSQL but slower than Umbra.
Both SQL variants (SQL-92/SQL + Arrays) perform better with an increasing number of tuples per iteration.
A small number of input tuples corresponds to a small batch size, leading to a small number of tuples used during one recursive step.
This thwarts database systems as they excel in batched processing.

\subsubsection{Memory Consumption}
The downside of the relational representation is its high memory consumption for matrix multiplication.
As it is not using any compression technique such as compressed sparse row, the matrices must be stored densely.
\autoref{fig:eval:irissql9210150} shows the memory consumption of Umbra for model training with inference on the whole Iris data set.
We performed ten iterations on a neural network with one hidden layer of size 20 (resp. 50).
Each iteration requires about 10 MiB (20 MiB) of main-memory, whereas its counterpart using arrays (cf.~\autoref{fig:eval:irisarray10150}) only requires 1 MiB (1.6 MiB).


\begin{table}
\centering
\sf\scriptsize
\caption{Matrix sizes exemplary for the Iris data set and a neural network with one hidden layer (20).}
\label{tab:sizes}
\begin{tabular}{|c|c|c|}\hline
Variable & \#Entries & Size in B \\\hline
$|x|$                                          & $150 \cdot 4 = 600$  & $600 \cdot 8$ \\
$|a_{xh}|=|l_{xh}|=|\delta_{xh}|$            & $150 \cdot 20= 3000$ & $ 3 \cdot 3000\cdot 8$ \\
$|a_{ho}|=|l_{ho}|=|\delta_{ho}|=|y_{ones}|$ & $150 \cdot 3 = 450$  & $ 4 \cdot 450 \cdot 8$ \\
$|w_{xh}|$                                     & $4   \cdot20 = 80$   & $ 80 \cdot 8$ \\
$|w_{ho}|$                                     & $20  \cdot3  = 60$   & $ 60 \cdot 8$ \\\hline\hline
Sum                                            &                      & $11540 \cdot 8$ \\\hline
\end{tabular}
\end{table}

\autoref{tab:sizes} shows the sizes of the involved tables:
The size of the weight matrices is independent of the data but depends on the size of the input/output vectors only (that are number of attributes, size of the hidden layer and number of categories).
The other matrices contain one row per input tuple.
For model inference, we need a total size of $(600+3000+450+450+80+20)\cdot 8B=4640 \cdot 8B = 36.25 KiB$.
Training requires to materialise the matrices for backpropagation ($90 KiB$).
As subsequent matrix operations on our array data type have not yet been optimised, we allocate a matrix for every single operation, which consumes additional memory.
Storing the matrices in a relational representation consumes a threefold of the memory as the indices each consume $8 B$.
As shown in Section~\ref{sec:relarray}, matrix multiplication consumes a multiple of the resulting matrix due to aggregation for summation.


\begin{figure}[tbp]
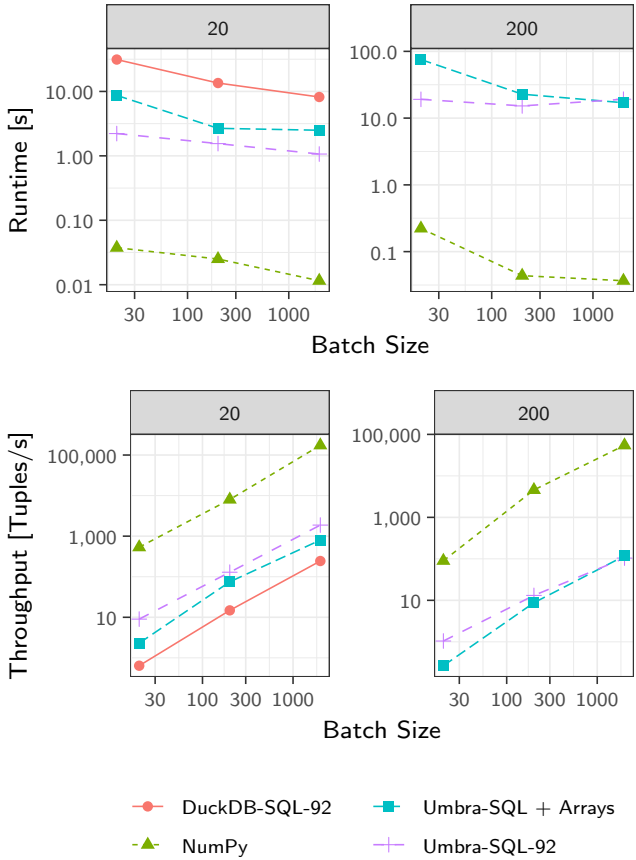

\sf
   \input{img/mnist.tex}
   \input{img/mnistthroughput.tex}
   \caption{Performance: Training one epoch with the MNIST data set with increasing batch size (one hidden layer size 20/200).}
   \label{fig:eval:mnist}
\end{figure}

\begin{figure}[tbp]
\sf
   \input{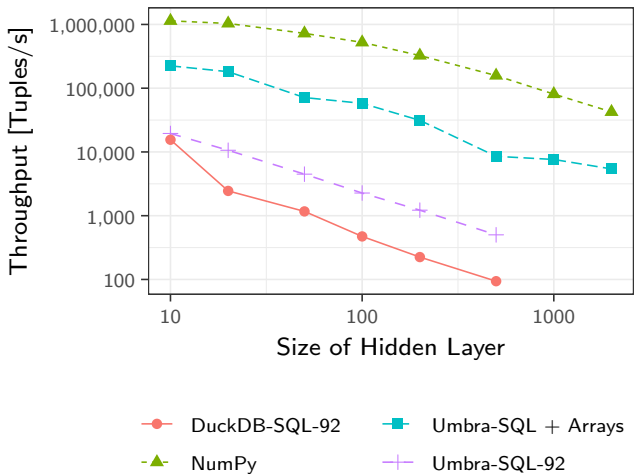}
   \caption{Throughput for model inference (MNIST data set) depending on the size of the hidden layer.}
   \label{fig:eval:mnistinference}
\end{figure}

\subsection{Image Classification}
The second benchmark simulates image classification based on the MNIST data set using a neural network with one hidden layer.
To investigate how database systems perform for model inference, we split the measurements for image classification in two parts for which we benchmark model inference separately.

\subsubsection{Throughput: Training}
We measure the runtime for training one epoch depending on the batch size.
As we can see in \autoref{fig:eval:mnist}, database systems perform better the bigger the batch size is.
With a higher batch size, the cost for aggregation into arrays is amortised and the SQL array data type outperforms the relational representation.
PostgreSQL was not terminating after a time-out of one day, DuckDB---when working in-memory only---exceeded the size of main-memory for the neural network with the larger hidden layer.
Only Umbra was capable of running all experiments.
To conclude, in-memory database systems are able to carry out matrix operations as required for neural networks.
Nevertheless, use-case-specific optimisations are needed to support smaller batch sizes.

\subsubsection{Throughput: Inference}
As training on small batches is time-expensive in SQL-92, we argue that inference alone is worthful inside a database system to avoid data extraction.
\autoref{fig:eval:mnistinference} shows the throughput just for model inference on the MNIST dataset in dependency on the size of the hidden layer.
Unfortunately, the relational representation performs the worst.
This underlies the need of a native integration for model inference.
A native integration can be either an operator such as the ModelJoin~\cite{DBLP:conf/edbt/KlabeHS23} or a data structure with support for matrix algebra.
The array data type in Umbra is such a data structure, although performing a factor of ten worse than the NumPy reference implementation.
On the other side, a database integration for inference eliminates data extraction, thus eliminating additional costs.
From a data engineer's perspective, one has to trade off between extraction and inference time.
Further, it must be acknowledged that the array data type has not yet been optimised for subsequent operations.
No optimisation leads to one separate call to the BLAS library for each operation, which decreases performance.
In the future, we plan the query optimiser to detect and combine subsequent matrix operation on an array data type to be executed as a single library call.

\subsubsection{Memory Consumption: Training}
\begin{figure}[tbp]
\sf
{
   \includegraphics[width=\linewidth]{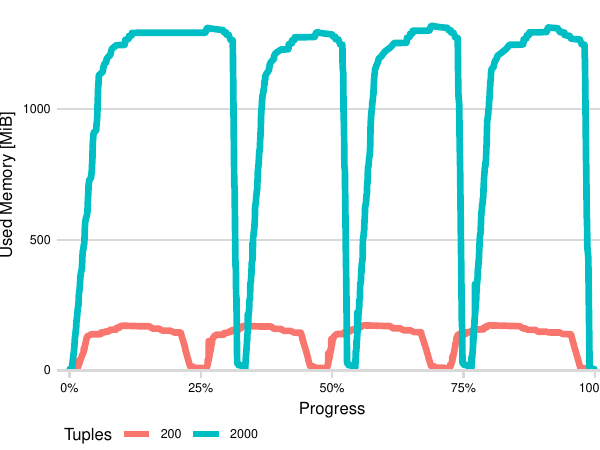}
   \subcaption{Hidden layer size 20.}
   \label{fig:eval:mnist_atts20}
   \includegraphics[width=\linewidth]{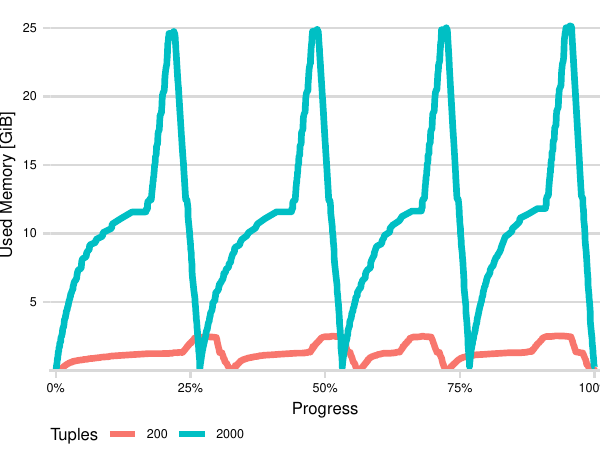}
   \subcaption{Hidden layer size 200.}
   \label{fig:eval:mnist_atts200}
}
   \caption{SQL-92: Memory consumption (batch size = 200/2000 Tuples).}
   \label{fig:eval:mnist_atts}
\end{figure}

\begin{figure}[tbp]
\sf
   \includegraphics[width=\linewidth]{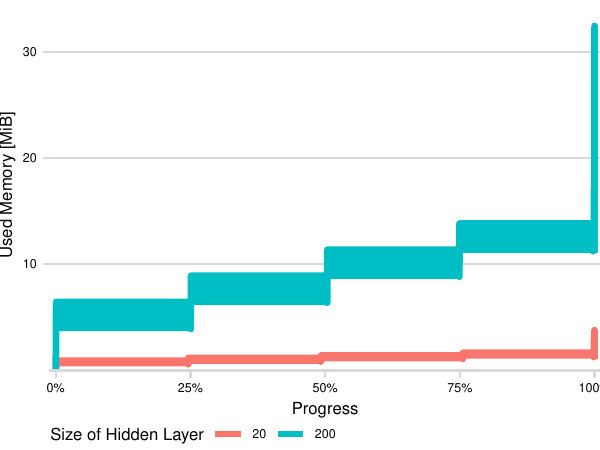}
   \caption{SQL + Arrays: Memory consumption (batch size = 2000 Tuples, hidden layer size 20/200).}
   \label{fig:eval:mnist_array2000}
\end{figure}

As the MNIST data set contains more columns (784) and labels (10) than the Iris data set, also the size of the intermediate tables grow within each iteration.
\autoref{fig:eval:mnist_atts} shows the memory consumption within Umbra for training (without inference) in a relational representation depending on the batch size for four iterations (not a complete epoch as the number of iterations would not match otherwise).
The higher the size of the hidden layer or the batch size, the higher the required memory for backpropagation gets.
The memory consumption reaches 25 GiB per iteration, whereas---when supporting arrays---only a fraction of the memory of its relational counterpart is required (less than 5~MiB per iteration, cf.~\autoref{fig:eval:mnist_array2000}).


\subsubsection{Memory Consumption: Inference}
\begin{figure}[tbp]
\sf
{
   \includegraphics[width=\linewidth]{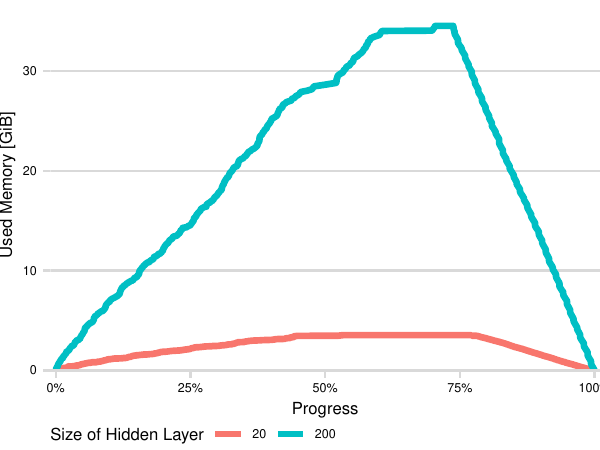}
   \subcaption{SQL-92}
   \label{fig:eval:inferencesql92}
   \includegraphics[width=\linewidth]{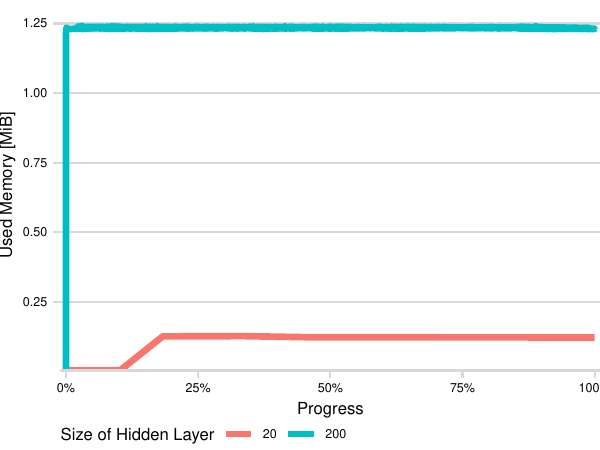}
   \subcaption{SQL + Arrays.}
   \label{fig:eval:inferencearray}
}
   \caption{Model Inference: Memory consumption (batch size = 6000 Tuples, hidden layer size 20/200).}
   \label{fig:eval:inferencemem}
\end{figure}

For inference only the forward pass of backpropagation is needed, thus less memory is required per tuple.
\autoref{fig:eval:inferencesql92} shows the memory usage of the relational representation for inference:
The two matrix multiplications that are involved for evaluating the model require each an aggregation in relational algebra and thus materialise the intermediate result.
This limits model inference by the number of input tuples and by the number of hidden layers.
To overcome these limitations, the input can be processed batchwise and stored to disk.
Further, another physical implementation of the aggregation operator that uses sorting instead of hashing would allow for continuous output as required for data streams as input.
An array data type reduces the memory consumption for model inference (cf.~\autoref{fig:eval:inferencearray}) allowing for more tuples to be labelled.

\section{Related Work}
\label{sec:relatedwork}
There are systems for machine learning, which enhance the application of models by integrating machine learning functionality into database systems, and machine learning for systems, which aims to replace components of systems with learned models.
This paper focuses on the first, as systems are fundamental for processing data and their enhancement is crucial for efficient training models and their inference~\cite{DBLP:conf/cikm/WangW22,DBLP:conf/vldb/ZhengTDK23,DBLP:conf/bncod/Mohr-DauratP21}.

\subsection{Database Systems for Machine Learning.}
For database systems completely taking over machine learning, the tasks can be either mapped to standardised SQL~\cite{DBLP:conf/edbt/SiksnysPNF21} as in Section~\ref{sec:implementationsql92} or database systems have to provide extensions~\cite{DBLP:journals/sigmod/JankovLYCZJG20,DBLP:journals/pvldb/HellersteinRSWFGNWFLK12,DBLP:conf/icml/TangDJYBJ23} as in Section~\ref{sec:implementationsqlarrays}.
Kläbe et~al.~\cite{DBLP:conf/edbt/KlabeHS23} focus on model inference, which means labelling unlabelled data based on a pre-trained model, for which they introduce the ModelJoin.
They use database systems for labelling only.
Paganelli et~al.~\cite{DBLP:journals/tkde/PaganelliSPIG23} also focus on model inference using case-when statements and one-hot-encoding to push prediction queries into the database system.
Yang~et~al.~\cite{DBLP:journals/pvldb/YangWHLLW22} focus on user defined functions for model inference and their optimisation.

Further studies propose to generate SQL queries for machine learning without modifying the underlying database system~\cite{DBLP:conf/icde/LuoGGPJ17,DBLP:journals/ojbd/MartenMDH19,DBLP:journals/corr/abs-2004-05366}.
Duta et~al.~\cite{DBLP:conf/cidr/DutaHG20} have proven how to translate procedural language extensions, i.e. PL/pgSQL, into recursive common table expressions (CTEs).
Their approach led to a performance increase through the elimination of costly procedural constructs in favour of parallelisable recursive SQL queries.
Based on this, Blacher et~al.~\cite{DBLP:journals/pacmmod/BlacherKSLLG23} use our relational representation for Einstein summation within code-generating database systems.
They are focusing on the performance for matrix multiplication whereas we also measured the memory consumption.
Other research focuses on database systems being part of the data preprocessing pipeline and accelerating data engineering.
Schleich et~al. \cite{DBLP:conf/sigmod/SchleichOK0N19} let the database system pre-aggregate data to solve regression tasks in SQL.
Yan~et~al.~\cite{DBLP:journals/pacmmod/YanLH23} accelerate prediction queries by detecting possible predicates to be executed within the database system beforehand.

\subsection{Systems Optimisation for Linear Algebra}
As the evaluation has shown that the memory consumption of matrices is limiting the amount of tuples to be processed, future extensions of database systems should incorporate research on matrix compression.
Elgohary et~al.~\cite{DBLP:journals/vldb/ElgoharyBHRR18} describe the trade-off between execution time and data size.
Baunsgaard et~al.~\cite{DBLP:journals/pacmmod/Baunsgaard023} compress matrices depending on the linear algebra required by workload.
Ferragina et~al.~\cite{DBLP:journals/pvldb/FerraginaMGKNST22} perform matrix-vector multiplication on a loss-less compressed matrix format based on compressed sparse row (CSR).
Other research eliminates any tabular representation by using matrices instead of tables~\cite{DBLP:journals/pvldb/HuangC22}.


\subsection{Machine Learning Systems.}
Different approaches try to unify the requirements for supporting machine learning algorithms with the advantages of database systems.
SystemDS~\cite{DBLP:conf/cidr/BoehmADGIKLPR20} and SystemML~\cite{DBLP:journals/pvldb/BoehmRHSEP18} are machine learning systems empowered by database technology such as index structures and a declarative language that lets the user define customised algorithms.
To support iterative algorithms on stored data within the database system, DB4ML~\cite{DBLP:conf/sigmod/Jasny0KRB20} defines an interface for iterative sub-transactions accessing stored tuples according to a data layout.
The work of Asada et al.~\cite{DBLP:journals/pvldb/AsadaFGGZGNBSI22} pursues the counter approach:
Instead of combining machine learning with database systems, they map database queries onto tensors.
In this way, they reuse machine learning frameworks such as PyTorch with its extensions for hardware accelerators~\cite{DBLP:journals/pvldb/HeNBSSPCCKI22}.
To specify the data layout for tensors instead of tuples, Schleich et~al.~\cite{DBLP:journals/corr/abs-2210-06267} developed a declarative tensor query language.
Other work try to avoid database systems and thus have to add functionality such as parallelisation for feature transformation~\cite{DBLP:journals/pvldb/PhaniEB22}.
In contrast to these works, we worked with database systems as the main component and with SQL as the principal query language.
We argue that database systems are already optimised for modern hardware but their optimiser needs tuning for machine learning workloads.

\section{Conclusion}
\label{sec:conclusion}
This paper has discussed and benchmarked building blocks for training neural networks in SQL.
In order to deduce the necessary SQL queries that represent matrix algebra for evaluating and training neural networks, we first discussed reverse mode automatic differentiation to reuse partial derivations.
The partial derivations formed the foundation for nested CTEs.
They were cached within a recursive CTE when deriving the weight matrices to compute the optimal weights.
Instead of CTEs for each variable, we presented an extension of SQL arrays for matrix algebra and the required data transformation steps.
In the evaluation, in-memory enhanced database systems, i.e. Umbra and DuckDB, showed better performance when processing data in larger batches, although they were still outperformed by NumPy in Python.
Furthermore, the available memory limited the number of tuples that could be processed for model training and inference within in-\-me\-mo\-ry systems.
As the relational representation materialised the data for training and inference, the extended array data type addressed this limitation by using less memory.
Although the set of matrix algebra extensions for array data types enabled the required operations, condensing subsequent calls would optimise memory usage further.

Future research is needed to optimise recursive CTEs for this use case and to automatically generate the presented queries.
As we were using recursion to imitate a procedural loop, the recursive CTE grew with each iteration.
This resulted in increased memory consumption per iteration, which limited the number of iterations and the model size.
To overcome these limitations, database optimisers should either eliminate intermediate results within the CTE or output intermediate results to free memory.

Hashing for aggregation hindered the system from continuously emitting tuples.
Thus, a sorting-based physical implementation would enable continuous output, which is desired for data streams as input.
Assuming these optimisations, the presented queries can be used to train more complex models with more weight variables.

To simplify the use of the relational representation, a transpiler that automatically generates the corresponding SQL queries from common array query languages such as NumPy or ArrayQL could be developed.
Such a transpiler could  offer additional features such as automatic differentiation for the generation of queries for model training and inference.

\balance
\bibliographystyle{spmpsci}      
\bibliography{paper}   

\begin{thebibliography}{10}
\providecommand{\url}[1]{{#1}}
\providecommand{\urlprefix}{URL }
\expandafter\ifx\csname urlstyle\endcsname\relax
  \providecommand{\doi}[1]{DOI~\discretionary{}{}{}#1}\else
  \providecommand{\doi}{DOI~\discretionary{}{}{}\begingroup
  \urlstyle{rm}\Url}\fi

\bibitem{DBLP:journals/pvldb/AsadaFGGZGNBSI22}
Asada, Y., Fu, V., Gandhi, A., Gemawat, A., Zhang, L., Gupta, V., Nosakhare,
  E., Banda, D., Sen, R., Interlandi, M.: Share the tensor tea: How databases
  can leverage the machine learning ecosystem.
\newblock Proc. {VLDB} Endow. \textbf{15}(12), 3598--3601 (2022)

\bibitem{DBLP:journals/pacmmod/Baunsgaard023}
Baunsgaard, S., Boehm, M.: {AWARE:} workload-aware, redundancy-exploiting
  linear algebra.
\newblock Proc. {ACM} Manag. Data \textbf{1}(1), 2:1--2:28 (2023)

\bibitem{DBLP:conf/cidr/BlacherGLKL22}
Blacher, M., Giesen, J., Laue, S., Klaus, J., Leis, V.: Machine learning,
  linear algebra, and more: Is {SQL} all you need?
\newblock In: {CIDR}. www.cidrdb.org (2022)

\bibitem{DBLP:journals/pacmmod/BlacherKSLLG23}
Blacher, M., Klaus, J., Staudt, C., Laue, S., Leis, V., Giesen, J.: Efficient
  and portable einstein summation in {SQL}.
\newblock Proc. {ACM} Manag. Data \textbf{1}(2), 121:1--121:19 (2023)

\bibitem{DBLP:conf/cidr/BoehmADGIKLPR20}
Boehm, M., Antonov, I., Baunsgaard, S., Dokter, M., Ginth{\"{o}}r, R.,
  Innerebner, K., Klezin, F., Lindstaedt, S.N., Phani, A., Rath, B., Reinwald,
  B., Siddiqui, S., Wrede, S.B.: Systemds: {A} declarative machine learning
  system for the end-to-end data science lifecycle.
\newblock In: {CIDR}. www.cidrdb.org (2020)

\bibitem{DBLP:journals/pvldb/BoehmRHSEP18}
Boehm, M., Reinwald, B., Hutchison, D., Sen, P., Evfimievski, A.V., Pansare,
  N.: On optimizing operator fusion plans for large-scale machine learning in
  systemml.
\newblock Proc. {VLDB} Endow. \textbf{11}(12), 1755--1768 (2018)

\bibitem{DBLP:journals/corr/abs-2207-14529}
Budach, L., Feuerpfeil, M., Ihde, N., Nathansen, A., Noack, N.S., Patzlaff, H.,
  Harmouch, H., Naumann, F.: The effects of data quality on ml-model
  performance.
\newblock CoRR \textbf{abs/2207.14529} (2022)

\bibitem{DBLP:journals/pvldb/ButtersteinMSBZ20}
Butterstein, D., Martin, D., Stolze, K., Beier, F., Zhong, J., Wang, L.:
  Replication at the speed of change - a fast, scalable replication solution
  for near real-time {HTAP} processing.
\newblock Proc. {VLDB} Endow. \textbf{13}(12), 3245--3257 (2020)

\bibitem{DBLP:conf/cvpr/CiresanMS12}
Ciresan, D.C., Meier, U., Schmidhuber, J.: Multi-column deep neural networks
  for image classification.
\newblock In: {CVPR}, pp. 3642--3649. {IEEE} Computer Society (2012)

\bibitem{DBLP:conf/tapp/Dietrich0G22}
Dietrich, B., M{\"{u}}ller, T., Grust, T.: Data provenance for recursive {SQL}
  queries.
\newblock In: TaPP, pp. 9:1--9:8. {ACM} (2022)

\bibitem{DBLP:journals/corr/abs-2004-05366}
Du, L.: In-machine-learning database: Reimagining deep learning with old-school
  {SQL}.
\newblock CoRR \textbf{abs/2004.05366} (2020)

\bibitem{DBLP:conf/cidr/DutaHG20}
Duta, C., Hirn, D., Grust, T.: Compiling {PL/SQL} away.
\newblock In: {CIDR}. www.cidrdb.org (2020)

\bibitem{DBLP:journals/vldb/ElgoharyBHRR18}
Elgohary, A., Boehm, M., Haas, P.J., Reiss, F.R., Reinwald, B.: Compressed
  linear algebra for large-scale machine learning.
\newblock {VLDB} J. \textbf{27}(5), 719--744 (2018)

\bibitem{DBLP:journals/pvldb/FerraginaMGKNST22}
Ferragina, P., Manzini, G., Gagie, T., K{\"{o}}ppl, D., Navarro, G., Striani,
  M., Tosoni, F.: Improving matrix-vector multiplication via lossless
  grammar-compressed matrices.
\newblock Proc. {VLDB} Endow. \textbf{15}(10), 2175--2187 (2022)

\bibitem{fisher1936use}
Fisher, R.A.: The use of multiple measurements in taxonomic problems.
\newblock Annals of eugenics \textbf{7}(2), 179--188 (1936)

\bibitem{DBLP:journals/pvldb/HeNBSSPCCKI22}
He, D., Nakandala, S.C., Banda, D., Sen, R., Saur, K., Park, K., Curino, C.,
  Camacho{-}Rodr{\'{\i}}guez, J., Karanasos, K., Interlandi, M.: Query
  processing on tensor computation runtimes.
\newblock Proc. {VLDB} Endow. \textbf{15}(11), 2811--2825 (2022)

\bibitem{DBLP:conf/debs/HeinrichLKB22}
Heinrich, R., Luthra, M., Kornmayer, H., Binnig, C.: Zero-shot cost models for
  distributed stream processing.
\newblock In: {DEBS}, pp. 85--90. {ACM} (2022)

\bibitem{DBLP:journals/pvldb/HellersteinRSWFGNWFLK12}
Hellerstein, J.M., R{\'{e}}, C., Schoppmann, F., Wang, D.Z., Fratkin, E.,
  Gorajek, A., Ng, K.S., Welton, C., Feng, X., Li, K., Kumar, A.: The madlib
  analytics library or {MAD} skills, the {SQL}.
\newblock Proc. {VLDB} Endow. \textbf{5}(12), 1700--1711 (2012)

\bibitem{DBLP:journals/pvldb/HuangC22}
Huang, Z., Chen, S.: Density-optimized intersection-free mapping and matrix
  multiplication for join-project operations.
\newblock Proc. {VLDB} Endow. \textbf{15}(10), 2244--2256 (2022)

\bibitem{DBLP:journals/sigmod/JankovLYCZJG20}
Jankov, D., Luo, S., Yuan, B., Cai, Z., Zou, J., Jermaine, C., Gao, Z.J.:
  Declarative recursive computation on an {RDBMS:} or, why you should use a
  database for distributed machine learning.
\newblock {SIGMOD} Rec. \textbf{49}(1), 43--50 (2020)

\bibitem{DBLP:conf/sigmod/Jasny0KRB20}
Jasny, M., Ziegler, T., Kraska, T., R{\"{o}}hm, U., Binnig, C.: {DB4ML} - an
  in-memory database kernel with machine learning support.
\newblock In: {SIGMOD} Conference, pp. 159--173. {ACM} (2020)

\bibitem{DBLP:conf/edbt/KlabeHS23}
Kl{\"{a}}be, S., Hagedorn, S., Sattler, K.: Exploration of approaches for
  in-database {ML}.
\newblock In: {EDBT}, pp. 311--323. OpenProceedings.org (2023)

\bibitem{DBLP:conf/icde/LuoGGPJ17}
Luo, S., Gao, Z.J., Gubanov, M.N., Perez, L.L., Jermaine, C.M.: Scalable linear
  algebra on a relational database system.
\newblock In: {ICDE}, pp. 523--534. {IEEE} Computer Society (2017)

\bibitem{DBLP:journals/pvldb/MaltryD22}
Maltry, M., Dittrich, J.: A critical analysis of recursive model indexes.
\newblock Proc. {VLDB} Endow. \textbf{15}(5), 1079--1091 (2022)

\bibitem{DBLP:journals/ojbd/MartenMDH19}
Marten, D., Meyer, H., Dietrich, D., Heuer, A.: Sparse and dense linear algebra
  for machine learning on parallel-rdbms using {SQL}.
\newblock Open J. Big Data \textbf{5}(1), 1--34 (2019)

\bibitem{DBLP:conf/edbt/MayLPMMCG15}
May, N., Lehner, W., P., S.H., Maheshwari, N., M{\"{u}}ller, C., Chowdhuri, S.,
  Goel, A.K.: {SAP} {HANA} - from relational {OLAP} database to big data
  infrastructure.
\newblock In: {EDBT}, pp. 581--592. OpenProceedings.org (2015)

\bibitem{DBLP:conf/icer/MiedemaAF21}
Miedema, D., Aivaloglou, E., Fletcher, G.: Identifying {SQL} misconceptions of
  novices: Findings from a think-aloud study.
\newblock In: {ICER}, pp. 355--367. {ACM} (2021)

\bibitem{DBLP:conf/bncod/Mohr-DauratP21}
Mohr{-}Daurat, H., Pirk, H.: Homoiconicity for end-to-end machine learning with
  {BOSS}.
\newblock In: {BICOD}, \emph{{CEUR} Workshop Proceedings}, vol. 3163, pp.
  46--49. CEUR-WS.org (2021)

\bibitem{MLPR}
Murray, I.: Machine learning and pattern recognition (mlpr): Backpropagation of
  derivatives (2017).
\newblock
  \urlprefix\url{https://www.inf.ed.ac.uk/teaching/courses/mlpr/2017/notes/w5a_backprop.pdf}

\bibitem{DBLP:journals/semweb/NathRPH22}
Nath, R.P.D., Romero, O., Pedersen, T.B., Hose, K.: High-level {ETL} for
  semantic data warehouses.
\newblock Semantic Web \textbf{13}(1), 85--132 (2022)

\bibitem{DBLP:conf/cidr/NeumannF20}
Neumann, T., Freitag, M.J.: Umbra: {A} disk-based system with in-memory
  performance.
\newblock In: {CIDR}. www.cidrdb.org (2020)

\bibitem{DBLP:conf/sigmod/OlteanuVZ22}
Olteanu, D., Vortmeier, N., Zivanovic, D.: Givens {QR} decomposition over
  relational databases.
\newblock In: {SIGMOD} Conference, pp. 1948--1961. {ACM} (2022)

\bibitem{DBLP:journals/tkde/PaganelliSPIG23}
Paganelli, M., Sottovia, P., Park, K., Interlandi, M., Guerra, F.: Pushing {ML}
  predictions into dbmss.
\newblock {IEEE} Trans. Knowl. Data Eng. \textbf{35}(10), 10,295--10,308 (2023)

\bibitem{DBLP:journals/pvldb/PhaniEB22}
Phani, A., Erlbacher, L., Boehm, M.: {UPLIFT:} parallelization strategies for
  feature transformations in machine learning workloads.
\newblock Proc. {VLDB} Endow. \textbf{15}(11), 2929--2938 (2022)

\bibitem{DBLP:conf/edbt/RaasveldtHMM18}
Raasveldt, M., Holanda, P., M{\"{u}}hleisen, H., Manegold, S.: Deep integration
  of machine learning into column stores.
\newblock In: {EDBT}, pp. 473--476. OpenProceedings.org (2018)

\bibitem{DBLP:conf/sigmod/RaasveldtM19}
Raasveldt, M., M{\"{u}}hleisen, H.: Duckdb: an embeddable analytical database.
\newblock In: {SIGMOD} Conference, pp. 1981--1984. {ACM} (2019)

\bibitem{DBLP:conf/sigmod/Renz-WielandGKM22}
Renz-Wieland, A., Gemulla, R., Kaoudi, Z., Markl, V.: Nups: {A} parameter
  server for machine learning with non-uniform parameter access.
\newblock In: {SIGMOD} Conference, pp. 481--495. {ACM} (2022)

\bibitem{DBLP:conf/ht/SalimzadehGHD22}
Salimzadeh, S., Gadiraju, U., Hauff, C., van Deursen, A.: Exploring the
  feasibility of crowd-powered decomposition of complex user questions in
  text-to-sql tasks.
\newblock In: {HT}, pp. 154--165. {ACM} (2022)

\bibitem{DBLP:conf/sigmod/SchleichOK0N19}
Schleich, M., Olteanu, D., Khamis, M.A., Ngo, H.Q., Nguyen, X.: A layered
  aggregate engine for analytics workloads.
\newblock In: {SIGMOD} Conference, pp. 1642--1659. {ACM} (2019)

\bibitem{DBLP:journals/corr/abs-2210-06267}
Schleich, M., Shaikhha, A., Suciu, D.: Optimizing tensor programs on flexible
  storage.
\newblock CoRR \textbf{abs/2210.06267} (2022)

\bibitem{DBLP:journals/pvldb/SchuhknechtPHS21}
Schuhknecht, F.M., Priesterroth, A., Henneberg, J., Salkhordeh, R.: Anyolap:
  Analytical processing of arbitrary data-intensive applications without {ETL}.
\newblock Proc. {VLDB} Endow. \textbf{14}(12), 2823--2826 (2021)

\bibitem{DBLP:conf/btw/Schule23}
Sch{\"{u}}le, M.E.: Recursive {SQL} and gpu-support for in-database machine
  learning.
\newblock In: {BTW}, \emph{{LNI}}, vol. {P-331}, p. 931. Gesellschaft f{\"{u}}r
  Informatik e.V. (2023)

\bibitem{DBLP:conf/ssdbm/SchuleGK021}
Sch{\"{u}}le, M.E., G{\"{o}}tz, T., Kemper, A., Neumann, T.: Arrayql for linear
  algebra within umbra.
\newblock In: {SSDBM}, pp. 193--196. {ACM} (2021)

\bibitem{DBLP:conf/edbt/SchuleGK022}
Sch{\"{u}}le, M.E., G{\"{o}}tz, T., Kemper, A., Neumann, T.: Arrayql
  integration into code-generating database systems.
\newblock In: {EDBT}, pp. 1:40--1:51. OpenProceedings.org (2022)

\bibitem{DBLP:conf/btw/SchuleK023}
Sch{\"{u}}le, M.E., Kemper, A., Neumann, T.: {NN2SQL:} let {SQL} think for
  neural networks.
\newblock In: {BTW}, \emph{{LNI}}, vol. {P-331}, pp. 183--194. Gesellschaft
  f{\"{u}}r Informatik e.V. (2023)

\bibitem{DBLP:conf/ssdbm/SchuleLSK0G21}
Sch{\"{u}}le, M.E., Lang, H., Springer, M., Kemper, A., Neumann, T.,
  G{\"{u}}nnemann, S.: In-database machine learning with {SQL} on gpus.
\newblock In: {SSDBM}, pp. 25--36. {ACM} (2021)

\bibitem{DBLP:conf/btw/SchulePK019}
Sch{\"{u}}le, M.E., Passing, L., Kemper, A., Neumann, T.: Ja-(zu-)sql:
  Evaluation einer sql-skriptsprache f{\"{u}}r hauptspeicherdatenbanksysteme.
\newblock In: {BTW}, \emph{{LNI}}, vol. {P-289}, pp. 107--126. Gesellschaft
  f{\"{u}}r Informatik, Bonn (2019)

\bibitem{DBLP:conf/sigmod/SchuleSBK021}
Sch{\"{u}}le, M.E., Schmei{\ss}er, J., Blum, T., Kemper, A., Neumann, T.:
  Tardisdb: Extending {SQL} to support versioning.
\newblock In: {SIGMOD} Conference, pp. 2775--2778. {ACM} (2021)

\bibitem{DBLP:conf/edbt/SiksnysPNF21}
Siksnys, L., Pedersen, T.B., Nielsen, T.D., Frazzetto, D.: Solvedb+: Sql-based
  prescriptive analytics.
\newblock In: {EDBT}, pp. 133--144. OpenProceedings.org (2021)

\bibitem{DBLP:conf/sigmod/StonebrakerR86}
Stonebraker, M., Rowe, L.A.: The design of postgres.
\newblock In: {SIGMOD} Conference, pp. 340--355. {ACM} Press (1986)

\bibitem{DBLP:conf/edbt/StorlK22}
St{\"{o}}rl, U., Klettke, M.: Darwin: {A} data platform for schema evolution
  management and data migration.
\newblock In: {EDBT/ICDT} Workshops, \emph{{CEUR} Workshop Proceedings}, vol.
  3135. CEUR-WS.org (2022)

\bibitem{DBLP:conf/icml/TangDJYBJ23}
Tang, Y., Ding, Z., Jankov, D., Yuan, B., Bourgeois, D., Jermaine, C.:
  Auto-differentiation of relational computations for very large scale machine
  learning.
\newblock In: {ICML}, \emph{Proceedings of Machine Learning Research}, vol.
  202, pp. 33,581--33,598. {PMLR} (2023)

\bibitem{DBLP:conf/cikm/WangW22}
Wang, Y., Wang, D.Z.: Extensible database simulator for fast prototyping
  in-database algorithms.
\newblock In: {CIKM}, pp. 5029--5033. {ACM} (2022)

\bibitem{DBLP:conf/edbt/WenigP22}
Wenig, P., Papenbrock, T.: Datagossip: {A} data exchange extension for
  distributed machine learning algorithms.
\newblock In: {EDBT}, pp. 2:373--2:377. OpenProceedings.org (2022)

\bibitem{DBLP:conf/sigmod/WieseH21}
Wiese, L., H{\"{o}}ltje, D.: Nncompare: a framework for dataset selection, data
  augmentation and comparison of different neural networks for medical image
  analysis.
\newblock In: DEEM@SIGMOD, pp. 6:1--6:7. {ACM} (2021)

\bibitem{DBLP:journals/pvldb/WingerathGR20}
Wingerath, W., Gessert, F., Ritter, N.: Invalidb: Scalable push-based real-time
  queries on top of pull-based databases (extended).
\newblock Proc. {VLDB} Endow. \textbf{13}(12), 3032--3045 (2020)

\bibitem{DBLP:journals/pacmmod/YanLH23}
Yan, C., Lin, Y., He, Y.: Predicate pushdown for data science pipelines.
\newblock Proc. {ACM} Manag. Data \textbf{1}(2), 136:1--136:28 (2023)

\bibitem{DBLP:journals/pvldb/YangWHLLW22}
Yang, Z., Wang, Z., Huang, Y., Lu, Y., Li, C., Wang, X.S.: Optimizing machine
  learning inference queries with correlative proxy models.
\newblock Proc. {VLDB} Endow. \textbf{15}(10), 2032--2044 (2022)

\bibitem{DBLP:conf/vldb/ZhengTDK23}
Zheng, Y., Tian, Y., D'silva, J.V., Kemme, B.: Dbmlsched: Scheduling
  in-database machine learning jobs.
\newblock In: {VLDB} Workshops, \emph{{CEUR} Workshop Proceedings}, vol. 3462.
  CEUR-WS.org (2023)

\end{thebibliography}

\end{document}